\documentclass[apjl]{emulateapj}

\usepackage[bottom]{footmisc}

\shorttitle{The electrical current density vector in the inner penumbra of a Sunspot}
\shortauthors{Puschmann et al.}

\begin{document}


\title{The electrical current density vector in the inner penumbra of a Sunspot}



\author{K. G. Puschmann\altaffilmark{1,2}, B. Ruiz Cobo\altaffilmark{1,2} and V. Mart\'\i nez Pillet\altaffilmark{1}}

\affil{(1) Instituto de Astrof\'isica de Canarias (IAC), E-38200 La Laguna, Tenerife, Spain\\
(2) Departamento de Astrof\'isica, Universidad de La Laguna (ULL), E-38205 La Laguna, Tenerife, Spain}
\email{kgp@iac.es, brc@iac.es, vmp@.iac.es}






\begin{abstract}
We determine the entire electrical current density vector in a geometrical 3D volume of the inner penumbra of a sunspot from an inversion of spectropolarimetric data obtained with Hinode/SP. Significant currents are seen to wrap around the hotter, more elevated regions with lower and more horizontal magnetic field that harbor strong upflows and radial outflows (the intraspines). The horizontal component of the current density vector is 3-4 times larger than the vertical; nearly all previous studies only obtain the vertical component $J_{\rm z}$ and thus strongly underestimate the current density. The current density $\vec J$ and the magnetic field $\vec B$ form an angle of about 20$^{\circ}$. The plasma $\beta$ at the 0\,km level is larger than 1 in the intraspines and is one order of magnitude lower in the background component of the penumbra (spines). At the 200\,km level, the plasma $\beta$ is below 0.3 nearly everywhere. The plasma $\beta$ surface as well as the surface optical depth unity are very corrugated. At the borders of intraspines and inside, $\vec B$ is not force-free at deeper layers and nearly force free at the top layers. The magnetic field of the spines is close to being potential everywhere. The dissipated ohmic energy is five orders of magnitudes smaller than the solar energy flux and thus negligible for the energy balance of the penumbra.
\end{abstract}


\section{Introduction}
The study of the stability or dynamics of penumbral filaments requires the knowledge of the Lorentz force, i.e. an accurate determination of the electrical current density vector $\vec J$ and the magnetic field vector $\vec B$. The estimation of the energy budget dissipated by electrical currents obviously requires the determination of $\vec J$. However, it is not trivial to reliably derive the electrical currents from observational data. Previous attempts were aimed almost exclusively at the determination of only the vertical component $J_{\rm z}$ with different degree of sophistication in the analysis of the data. The bulk of such studies was based on magnetograms \citep[][]{deloachetal84, hagyard88, hofmannetal88, hofmannetal89, canfieldetal92, delaBeaujardiere93, lekaetal93, metcalfetal94, vandriel94, wangetal94, zhangwang94, garydemoulin95, lietal97,  gaoetal08}. More accurate estimates  of $J_{\rm z}$ stem from state-of-art spectropolarimetric observations and the application of inversion techniques. In case of Milne-Eddington (ME) inversions \citep [e.g.][]{skumanichlites87, laggetal04}, $\vec B$ is obtained at an average optical depth. Among the recent works calculating  $J_{\rm z}$ under ME approximation one finds \citet{shimizuetal09}, \citet{venkatakrishnantiwari09}, and \citet{lietal09}. An alternative determination of $J_{\rm z}$ is obtained by \citet{balthasar06}, \citet{jucaketal06}, and \citet{balthasargoemoery08} employing SIR \citep[Stokes Inversion based on Response functions,][]{ruizcobodeltoro92}, which delivers the  stratification of $\vec B$ in an optical depth scale. 

There have been several attempts to obtain $J_{\rm hor}$, the modulus of the horizontal component of $\vec J$, imposing approximations to the magnetic field distribution: \citet{jietal03} and \citet{georgoulislabonte04} obtain a lower limit of $J_{\rm hor}$ assuming a field-free configuration; \citet{pevtsovperegud90} derive $J_{\rm hor}$ imposing cylindrical symmetry of $\vec B$ in a sunspot. Without making any hypothesis on $\vec B$, the determination of the three components of $\vec J$ requires the knowledge of the entire magnetic field vector in a geometrical 3D volume, i.e. the determination of a {\em geometrical} height scale is mandatory. While commonly in quiet Sun the transformation from an optical depth scale to  geometrical heights is done by assuming hydrostatic equilibrium \citep[see, e.g.][]{pus05}, this is not justified in the magnetized penumbra. In this case the force balance must include magnetic forces which requires to calculate the horizontal and vertical spatial derivatives of $\vec B$. A first attempt to empirically derive 3D vector currents was done by \citet{socasnavarro05}, who determined a geometrical height scale \citep[following ][]{san05} by imposing equal total pressure between adjacent pixels, although neglecting the magnetic tension in the Lorentz force. 

For an accurate determination of $\vec J$ in the present paper, we take advantage of the 3D geometrical model of a section of the inner penumbra of a sunspot  described in \citet{puschmannetal10} [hereafter, Paper I]. We use observations of the active region AR 10953 near solar disk center obtained on 1$^{\rm st}$ of May 2007 with the  Hinode/SP. The inner, centerside, penumbral area under study was located at an heliocentric angle $\theta$\,=\,4.63$^{\circ}$. To derive the physical parameters of the solar atmosphere as a function of continuum optical depth, the SIR inversion code was applied on the data set. The 3D geometrical model was derived by means of a genetic algorithm that minimized the divergence of the magnetic field vector and the deviations from static equilibrium considering pressure gradients, gravity and the Lorentz force. For a detailed description we refer to Paper I.

\section{Results and discussion}

 \begin{figure}
\begin{center}   
\includegraphics[scale=.7]{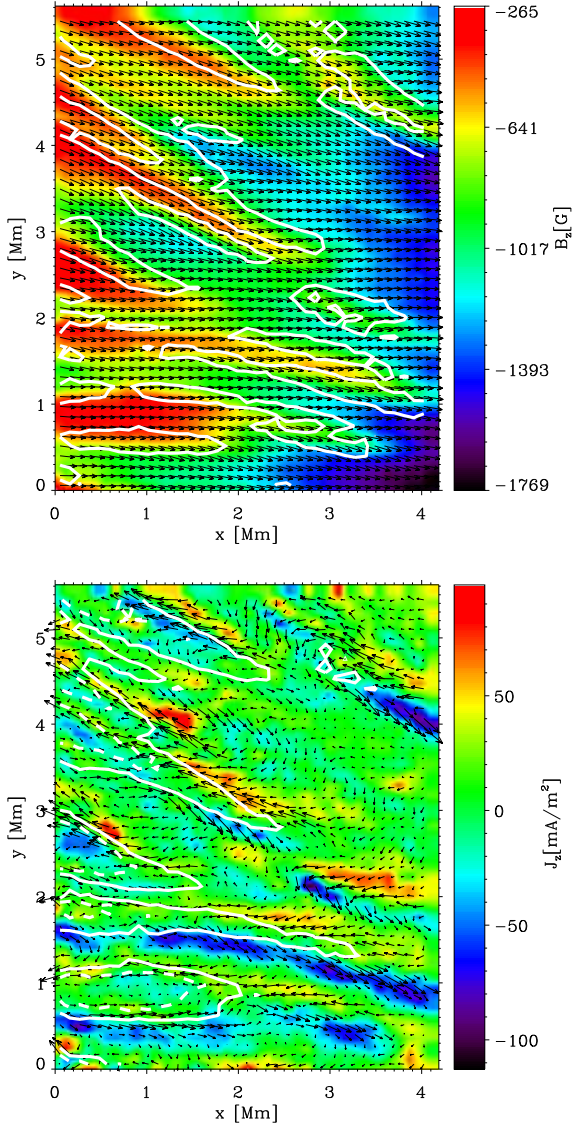} 
\end{center}
      \caption{Upper panel: $B_{\rm z}$ (in color) and $\vec{B}_{\rm hor}$ (arrows) at $z$\,=\,200\,km. The size of the arrows ranges from 1220 to 1900\,G. Contour lines enclose areas with $J$\,$>$\,120\,mA/m$^{2}$. Lower panel: $J_{\rm z}$ (in color) and $\vec{J}_{\rm hor}$ (arrows) at $z$\,=\,200\,km. The size of the arrows ranges from 1 to 365\,mA/m$^{2}$. Contour lines correspond to $B_{\rm z}$ equal to -650\,G (solid lines) and to -450\,G (dashed lines).}
         \label{Fig1}
   \end{figure}

The current density vector $\vec J$\,=\,$(\nabla\times\vec{B})/\mu_{0}$ has been calculated for each pixel in the FOV analyzed in Paper I at geometrical height layers between 0 and 200 km, i.e. in a volume of (4.2\,Mm x 5.6\,Mm x 0.2\,Mm) in the inner penumbra of a sunspot placed close to the disk center.
We decompose $\vec{B}=B_{\rm z}\vec{e_{\rm z}}+\vec{B}_{\rm hor}$ where  $\vec{e_{\rm z}}$ denote the unit vector in vertical direction. In a similar way, we write $\vec{J}=J_z\vec{e_{\rm z}}+\vec{J}_{\rm hor}$. Through the paper $B$, $B_{\rm hor}$, $J$ and $J_{\rm hor}$ will denote the modulus of the corresponding vectors. The upper panel of Fig.~\ref{Fig1} shows  $B_{\rm z}$ (colored background) and $\vec{B}_{\rm hor}$ (black arrows) at the top layer (200\,km). Areas in red color correspond to regions with lower and more horizontal magnetic field \citep[intraspines,][]{litesetal93} which are hotter, elevated and harbor strong upflows and radial outflows (see Paper I). These regions can be interpreted as the embedded nearly horizontal flux tubes of the uncombed scenario \citep[see e.g.][and references therein]{solankimontavon93, sch98a, sch98b, marpill00, borrerosolanki10}. We will call spines to regions with a more vertical and intense magnetic field (they correspond to the background component in the uncombed scenario). White contour lines enclose areas with significant $J$ values (larger than 120 mA/m$^{2}$), located predominantly along the borders of the intraspines. The lower panel of Fig.~\ref{Fig1} shows $J_{\rm z}$ (colored background) and  $\vec{J}_{\rm hor}$ (black arrows). White contour lines correspond to $B_{\rm z}$ values equal to -450 (solid) and -650 G (dashed), respectively. The electrical currents wrap around the nearly horizontal flux tubes (intraspines): For the majority of the intraspines, $J_{\rm z}$ shows positive values at the upper (larger Y-coordinate) borders of the filaments and negative values at the lower borders. The orientation of the black arrows indicates that the currents circumvent the flux tubes in most cases. 

   \begin{figure*}
\begin{center}
   \includegraphics[scale=.55]{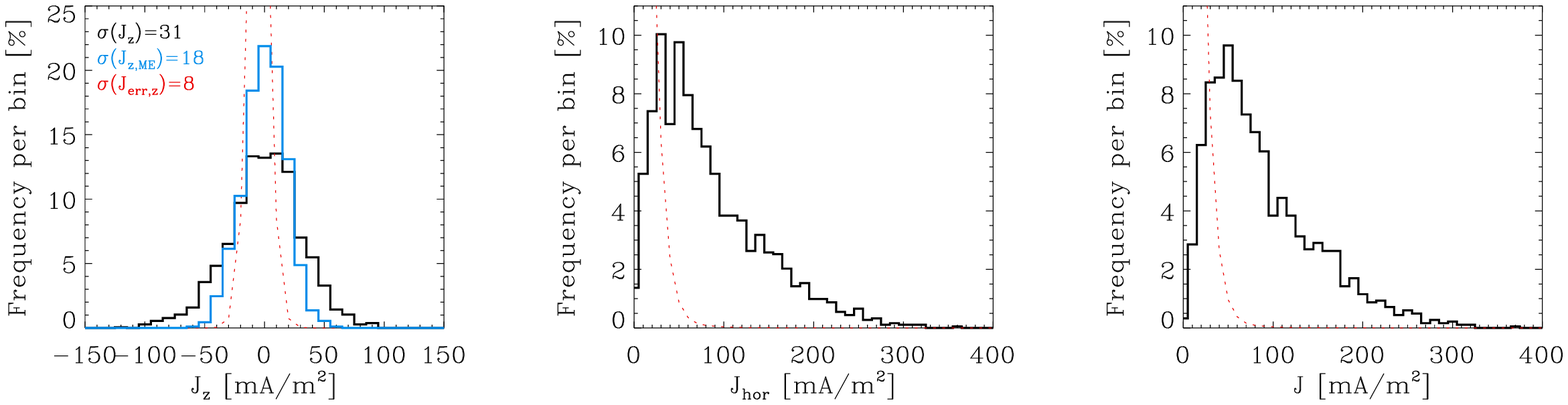}
\end{center}
     \caption{Left panel: Histogram of $J_{\rm z}$ at 200 km (black) and histogram of $J_{\rm z}$ retrieved from a Milne-Eddington inversion (blue). Middle and right panel: histograms of $J_{\rm hor}$ and $J$ evaluated at the same height layer. Red lines represent the corresponding error distributions.}
         \label{Fig2}
   \end{figure*}

\begin{table}
\begin{center}
\label{table1}
\caption{Mean values of the moduli of the vertical and horizontal component of $\vec J$ as well as the mean value of $J$ at three different height layers.}

\begin{tabular}{rrrr}
\hline
$z$  & $<\mid{J_{\rm z}}\mid>$ & $<{J_{\rm hor}}>$ & $<J>$ \\
{[km]} & [mA/m$^{2}$] & [mA/m$^{2}$] & [mA/m$^{2}$] \\
\hline
    0  & 30\,$\pm$\,20 & 129\,$\pm$\,50  & 135\,$\pm$\,59 \\
\hline
    100  & 26\,$\pm$\,\,\,\,9 & 103\,$\pm$\,36  & 109\,$\pm$\,39 \\
\hline
    200  & 25\,$\pm$\,\,\,\,8 & 88\,$\pm$\,33  & 93\,$\pm$\,35 \\
\hline
    ME  & 15\,\,\,\,\,\,    &   & \\
\hline
\end{tabular}
\end{center}
\end{table}

In the left panel of Fig.~\ref{Fig2} we plot the histogram of $J_{\rm z}$ evaluated at the top layer (200 km) in black.
To check the significance of the results, we evaluate a simulated electrical current density distribution $\vec{J}_{\rm err}$. The random vector field $\vec{J}_{\rm err}$ deviates from zero only by a Gaussian noise with a $\sigma$ equal to the estimated error at each pixel for each component of $\vec J$; $\sigma$ was obtained from an error propagation of the uncertainties of $\vec B$. The histogram of $J_{\rm{err,z}}$ is represented in red. Frequently, the $z$ component of the current density is evaluated from the results of ME inversions; to determine the reliability of such results we performed a ME analysis of our data set, and we calculated the vertical current density $J_{\rm{z,ME}}$. The histogram of $J_{\rm{z,ME}}$ is plotted in blue in the left panel of Fig.~\ref{Fig2}. Note that the $J_{\rm z}$ calculated from our 3D geometrical model is significantly larger than the $J_{\rm{z,ME}}$, however both are clearly above the calculated uncertainties. The ME inversion delivers the magnetic field just at an average optical depth around $\log\tau$\,=\,-1.5 \citep{ruizcobodeltoro94}. As we have seen before, larger electrical currents appear at the borders of structures with lower magnetic field and smaller optical depth. Consequently, the ME inversion probes higher layers above these structures resulting in a smaller derivative of the magnetic field as the one obtained from the 3D geometrical model. In the middle and right panel of Fig.~\ref{Fig2}, we present the histograms of $J_{\rm hor}$ and $J$, respectively. In red we show again the estimated uncertainties. As mentioned before, in most observational studies of electrical currents in solar active regions only the vertical component has been measured. We find $J_{\rm hor}$ to be about four times larger than $J_{\rm z}$. Note that the horizontal component is clearly predominant, with a distribution practically equal to the one of $J$. Works estimating $J_{\rm hor}$ by making certain hypotheses on the $\vec B$ configuration reach similar results: \citet{pevtsovperegud90} and \citet{georgoulislabonte04} found $J_{\rm hor}$\,$\sim$\,2-3 times larger than $J_{\rm z}$, while \citet{jietal03} report on $J_{\rm hor}$ values of about one or two orders larger than $J_{\rm z}$. All studies estimating only $J_{\rm z}$ strongly underestimate $\vec J$ present in sunspot penumbrae. 

Table 1 summarizes the values of $<\mid{J_{\rm z}}\mid>$, $<{J_{\rm hor}}>$, and $<J>$ at three different heights. The errors given in the table have been calculated by an error propagation from the distribution of $\vec{J}_{\rm err}$ described above. The electrical current density decreases with height and at all layers the horizontal component is on average about four times larger than the vertical one. While $<{J_{\rm hor}}>$ decreases by 32\% between the 0 and 200\,km level, $<\mid{J_{\rm z}}\mid>$ shows a small decrease with height of about 17\%. The large uncertainties at the 0\,km level stem from the corresponding uncertainties of the determination of the magnetic field due to the low sensitivity of the visible lines used in this study.

   \begin{figure}
\begin{center}
   \includegraphics[scale=.40]{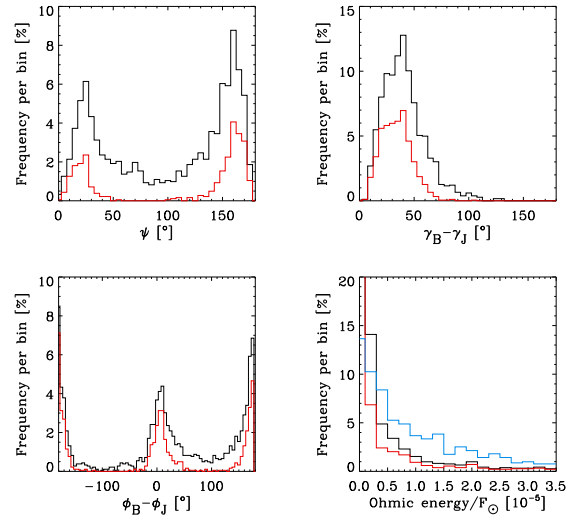} 
\end{center}
      \caption{Upper left panel: Histogram of the angle $\psi$ between $\vec J$ and $\vec B$ at $z$\,=\,0\,km. Upper right panel: Histogram of the difference between the inclination from the vertical of $\vec B$ and $\vec J$ at $z$\,=\,0\,km. Lower left panel: The same for the difference in azimuth. Lower right panel: Energy dissipated by electrical currents in units of $10^{-5}$ solar flux (between 0 and 200\,km, black; between -225 and 200\,km, blue). In all panels the red line corresponds to angles and energies evaluated for those pixels with $J(z=200)$\,$>$\,120\,mA/m$^{2}$.}
         \label{Fig3}
   \end{figure}

 The upper left panel of Fig.~\ref{Fig3} shows the histogram of the angle $\psi$ formed by $\vec B$ and $\vec J$ at the 0\,km height layer (black). In a force-free configuration, $\vec B$ and $\vec J$ are parallel to each other. In our volume $\psi$ is of about 20$^{\circ}$. Since the uncertainty of $\psi$ is large for pixels with small $J$, we show also the histogram of $\psi$ for pixels with $J(z=200)$\,$>$\,120\,mA/m$^{2}$ (red line). The difference in the orientation of $\vec B$ and $\vec J$ is mainly due to a difference in their inclination from the vertical $\gamma$ and not due to a difference in their azimuth $\phi$ as the upper right and lower left panel of Fig.~\ref{Fig3} indicate: Both vector fields are basically in the same vertical plane. Consequently, the horizontal component of the Lorentz force must be larger than the vertical one. In fact $<(\vec{J}\times\vec{B})_{\rm hor}>$ amounts to 1.03\,millidynes/cm$^{3}$, whereas $<(\vec{J}\times\vec{B})_{\rm z}>$ shows a value of 0.61\,millidynes/cm$^{3}$ only, both evaluated at $z$\,=\,0\,km for pixels with $J(z=200)$\,$>$\,120\,mA/m$^{2}$. This is in agreement with \citet{pevtsovperegud90} who also found the horizontal component of the Lorentz force being larger than the vertical one.

The lower right panel of Fig.~\ref{Fig3} shows the histograms of the energy dissipated by the electrical currents (ohmic energy) in the region under study  integrated between the 0 and 200\,km level (black). The  ohmic energy has been calculated using the longitudinal electrical conductivity following \citet{kopeckykuklin69} and is negligible in the inner sunspot penumbra compared with the solar flux.  The energy is dissipated only at the borders of the horizontal tubes, where $J$ reaches significant values (red line). In Paper I we found that the main contribution to the energy flux carried by the ascending mass (convective energy) stems from layers below -75\,km. The resulting convective energy flux, integrated from -225 to 200\,km, reaches values of up to 78\% of the solar flux, and thus would be sufficient to explain the observed penumbral brightness. However, this result has to be taken with care since the physical parameters at layers below -75\,km entering this calculation result from extrapolations. If all the layers between -225 and 200\,km are included in the calculation of the ohmic energy, the resulting values (well below 10$^{-4}$\,F$_{\odot}$) are still negligible for the energy balance (blue line in Fig.~\ref{Fig3}).

  \begin{figure}
\begin{center}
    \includegraphics[scale=.60]{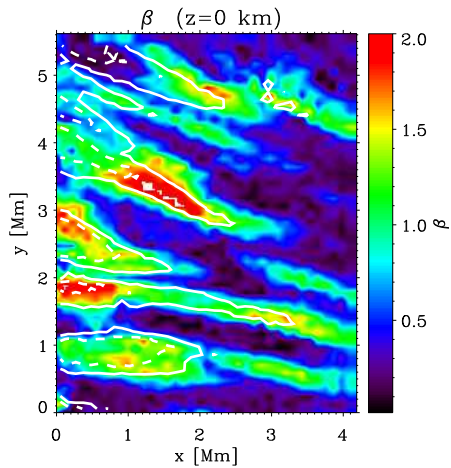}
\end{center}
       \caption{Map of the plasma $\beta$ at z\,=\,0\,km. Contour lines correspond to $B_{\rm z}$ equal to -650\,G (solid lines) and to -450\,G (dashed lines).}
          \label{Fig4}
    \end{figure}

The plasma $\beta$, defined as the ratio between the gas pressure and the magnetic pressure, is an important parameter to study the relevance of the Lorentz force in the dynamics of sunspot penumbrae. Fig.~\ref{Fig4} shows the plasma $\beta$ at a height of 0\,km. In our inner penumbral area, the $\beta$\,=\,1 surface is strongly corrugated. Note that in the intraspines (areas enclosed by white contour lines) the plasma $\beta$ is clearly larger than 1, being one order of magnitude lower in the spines. At the top layer (200\,km) not presented here, the aspect is similar, although all values have decreased by a factor $\sim$\,5.

\section{Conclusions}
The application of a genetic algorithm on the optical depth model retrieved from a SIR inversion of spectropolarimetric Hinode/SP data allowed us to construct a 3D geometrical model of a section of the inner penumbra of a sunspot (see Paper I). The resulting model has, by construction, a minimal divergence of $\vec B$ and a minimal deviation of the static equilibrium. This allows us for the first time the accurate determination of the entire 3D electrical current density vector. At the 0\farcs32 resolution of the Hinode/SP, the electrical current density appears to be spatially resolved in the penumbra: Significant $J$ values (larger than 120 mA/m$^{2}$) are located at the borders of the intraspines. The electrical currents seem to wrap around the nearly horizontal flux tubes.

{We find an horizontal component of $\vec J$ about four times larger than the vertical one. This is in agreement with earlier works estimating $J_{\rm hor}$ by assuming simplifying hypotheses on the $\vec B$ distribution. All works$^{1}$ only evaluating $J_{\rm z}$ clearly underestimate $\vec J$.

The magnetic field at lower layers is not force-free at the borders of the intraspines: There, the angle formed by $\vec B$ and $\vec J$ is about 20$^{\circ}$. The difference in the orientation of $\vec B$ and $\vec J$ is mainly due to a different inclination with respect to the vertical of both vectors, being nearly negligible the difference in azimuth, leading to a dominant horizontal component of the Lorentz force, directed towards the central axis of the intraspines, something that helps them to maintain the internal force balance. 
At the 0\,km level, the plasma $\beta$ is strongly corrugated, being larger than 1 at the borders and inside the intraspines and one order of magnitude lower inside the spines. In the latter, the Lorentz force can hardly be balanced by the pressure gradient or weight of the material: consequently the magnetic field configuration of the spines in the inner penumbra must be close to a force-free configuration.
Furthermore, as in these areas $J$ is relatively small and $B$ is large, the field must be closer to being potential. Following the same reasoning, our results show that at the highest layers the plasma $\beta$ is nearly everywhere below 0.3 and thus the field must be closer to a force-free configuration almost everywhere and to a potential one in the spines.

The dissipated ohmic energy is clearly negligible being 5 orders of magnitudes smaller than the solar flux. 

\acknowledgments
This work has been supported by the Spanish Ministerio de Ciencia e Innovaci\'on through projects ESP 2006-13030-C06-01, AYA2007-65602, AYA2009-14105-C06-03, AYA2007-63881, and the European Commission through the SOLAIRE Network (MTRN-CT-2006-035484). We thank C. Beck for fruitfull discussion.


\clearpage

\end{document}